\documentclass[aps,prb,twocolumn]{revtex4}
\usepackage{graphicx}
\begin{document}
\title{Multiple and virtual photon processes in radiation-induced magnetoresistance
oscillations in two-dimensional electron systems}
\author{X. L. Lei and S. Y. Liu}
\affiliation{Department of Physics, Shanghai Jiaotong University,
1954 Huashan Road, Shanghai 200030, China}

\begin{abstract}
Recently discovered new structures and zero-resistance states 
outside the well-known oscillations are demonstrated to arise from 
multiphoton assisted processes, by a detailed analysis of microwave 
photoresistance in two-dimensional electron systems under enhanced radiation. 
The concomitant resistance dropping and the peak narrowing observed in the experiments
are also reproduced. We show that the 
radiation-induced suppression of
average resistance comes from virtual photon effect 
and exists throughout the whole magnetic field range.
\end{abstract}


\maketitle

The recent interest in radiation related magneto-transport in two-dimensional (2D) 
electron gas (EG) 
has been stimulated by the discovery of microwave induced magnetorersistance oscillations (MIMOs) 
and zero-resistance states (ZRS) in ultra-high mobility systems at low temperatures.\cite{Zud01,Ye,Mani,Zud03}
Tremendous experimental\cite{Yang,Dor03,Mani04,Will,Kovalev,Mani-apl,Zud04,Stud,Dor05,Zud05} 
and theoretical\cite{Ryz,Ryz86,Anderson,Koul,Andreev,Durst,Xie,Lei03,Dmitriev,Ryz05199,
Vav,Mikh,Dietel,Torres,Dmitriev04,Inar,Lei05,Ryz05,Joas05,Ng} efforts have been devoted to study 
this exciting phenomenon and a general understanding of it has been reached.
MIMOs emerge as the magnetoresistance  $R_{xx}$ of the 2D system subject to a microwave radiation 
of frequency $f=\omega/2\pi$, exhibiting periodic oscillation  
as a function of the inverse magnetic field $1/B$.
It features the periodical appearance of peak-valley pairs 
around $\omega/\omega_c=1,2,3,4,\cdots$, i.e. 
a maximum at $\omega/\omega_c=j-\delta_j^-$ and 
a minimum at $\omega/\omega_c=j+\delta_j^+$, with $j=1,2,3,4,\cdots$ and 
$0<\delta_j^{\pm}\leq 1/4$.
Here $\omega_c$ is the cyclotron frequency and $\omega/\omega_c=j$ are  
the node points of the oscillation.
With increasing the radiation intensity 
the minimum value of $R_{xx}$ drops down 
until a vanishing resistance is measured, i.e. ZRS. 

In addition to these well-established features, 
a clear secondary structure between the first and second main peaks was
observed experimentally\cite{Mani,Zud03} and predicted theoretically\cite{Lei03} 
with reference to two-photon process.\cite{Lei03,Zud04} 
A distinct minimum and/or maximum outside the first main peak was also 
observed experimentally at 47.5, 52 and 58\,GHz,\cite{Mani,Mani04} 
at 45 and 35\,GHz\cite{Zud01,Zud03}, and at 30\,GHz\cite{Dor03}
which was referred to two-photon\cite{Mani04,Zud04} or the second harmonic
effect.\cite{Dor03} 
Theoretically a similar structure due to two-photon
process was anticipated at 60 and 40\,GHz,\cite{Lei03} followed by further
experimental observations at lower frequency (10-50\,GHz).\cite{Will,Zud04,Dor05} 
A more prominent experimental result was recently reported by Zudov {\it et al} 
at 27\,GHz,\cite{Zud05} where, with intensified microwave radiation the above-mentioned minima
evolve into ZRS.
Despite the consensus on the existence of these structures the assignment 
of their locations has so far been different among different groups.\cite{Mani04,Zud04}
Physically these peak-valley structures can be due to multi-photon
process,\cite{Lei03,Mani04,Zud04,Lei05,Zud05} or due to higher harmonics\cite{Dor03} 
of the radiation-induced high-frequency current.
To ascertain these structures as arising from multiphoton effect 
and to help identifying their positions, it is imperative
to reproduce them convincingly from a careful theoretical calculation 
extended to higher radiation intensity.

Another feature is the descent of the average 
dissipative resistance under microwave irradiation,
which was experimentally observed on the high magnetic field side 
where MIMO shows up relatively weak.\cite{Dor03,Mani-apl,Dor05,Zud05}
Theoretically, the magnetoresistance drop was also anticipated\cite{Ryz05,Lei05} 
and referred to virtual photon effect,\cite{Lei05} or 
to microwave-induced dynamic localization.\cite{Ryz05}
Nevertheless, the range and the degree of this resistance descent remain unclear,
and a unified theoretical treatment covering higher radiation intensity to demonstrate 
it together with multiphoton structures is desirable. 

In this Letter we report on a detailed analysis of  MIMOs with enhanced radiation 
intensity, based on a theoretical model which covers all orders of real and virtual
photon processes of the base frequency, but excludes higher harmonic current response.

The model considers that a dc electric field ${\bf E}_0$ and a high frequency (HF) field 
${\bf E}(t)\equiv{\bf E}_s \sin(\omega t)+{\bf E}_c\cos(\omega t)$ 
 are applied in a quasi-2D system consisting of $N_{\rm e}$ 
interacting electrons in a unit area of the $x$-$y$ plane, 
together with a magnetic field ${\bf B}=(0,0,B)$ along the $z$ direction.
The approach is based on the separation of the center-of-mass motion 
from the relative electron motion of the electrons and 
describes the transport state of a high-carrier-density many-electron system
under a radiation field in terms of a time-dependent electron drift velocity
oscillating with base radiation frequency, 
$
{\bf v}(t)={\bf v}_c \cos(\omega t)+{\bf v}_s \sin(\omega t)
$, and another part ${\bf v}_0$, describing the slowly varying 
electron drift motion,  
together with an electron temperature $T_{\rm e}$ characterizing
the electron heating.\cite{Lei85,Liu} In the case of ultra-clean electron gas
at low temperatures,
${\bf v}_0$ and $T_{\rm e}$ satisfy 
the following force- and energy-balance equations:\cite{Lei03,Lei05}
\begin{eqnarray}
&&m\frac{d{\bf v}_{0}}{dt}=e{\bf E}_{0}+ e ({\bf v}_0 \times {\bf B})+
\frac{{\bf F}_0}{N_{\rm e}},\label{eqv0}\\
&&N_{\rm e}{\bf E}_0\cdot {\bf v}_0+S_{\rm p}- W=0,
\label{eqne}
\end{eqnarray}
with ${\bf v}_c$ and ${\bf v}_s$ determined by
\begin{eqnarray}
-{m\omega}{\bf v}_{c}&=&{e{\bf E}_s}
+e({\bf v}_{s}\times
{\bf B}),\label{eqv1}\\
{m\omega}{\bf v}_{s}&=&{e{\bf E}_c}
+e({\bf v}_{c}
\times {\bf B}).\label{eqv2}
\end{eqnarray}
Here $e$ and $m$ are the electron charge and effective mass, 
\begin{equation}
{\bf F}_{0}=\sum_{{\bf q}_\|}\left| U({\bf q}_\|)\right| ^{2}
\sum_{n=-\infty }^{\infty }{\bf q}_\|{J}_{n}^{2}(\xi )
\Pi _{2}({\bf q}_\|,\omega_0-n\omega ) \label{eqf0}
\end{equation}
is the damping force of the moving center of mass, 
\begin{equation}
S_{\rm p}=\sum_{{\bf q}_\|}\left| U({\bf q}_\|)\right| ^{2}
\sum_{n=-\infty }^{\infty }n\omega J_{n}^{2}(\xi )
\Pi _{2}({\bf q}_\|,\omega_0-n\omega )
\label{eqsp}
\end{equation}
is the averaged rate of the electron energy absorption from the HF field, and 
\begin{equation}
W=\sum_{{\bf q}}\left| M({\bf q})\right|
^{2}\sum_{n=-\infty
}^{\infty } \Omega_{\bf q}J_{n}^{2}(\xi )\Lambda _{2}({\bf q},\omega_0+
\Omega _{{\bf q}}-n\omega )
 \label{eqw}
\end{equation}
is the average rate of the electron energy loss to the lattice. 
In the above equations, $J_n(\xi)$ is the Bessel function of order $n$,
$
\xi\equiv \sqrt{({\bf q}_\|\cdot {\bf v}_c)^2+
({\bf q}_\|\cdot {\bf v}_s)^2}/{\omega}
$;
$\omega_0\equiv {\bf q}_\|\cdot {\bf v}_0$,
$U({\bf q}_\|)$ and $M({\bf q})$ are effective impurity and phonon
scattering potentials, 
$\Pi_2({\bf q}_\|,\Omega)$ and
$\Lambda_2({\bf q},\Omega)=2\Pi_2({\bf q}_\|,\Omega)
[n(\Omega_{\bf q}/T)-n(\Omega/T_{\rm e})]
$ (with $n(x)\equiv 1/({\rm e}^x-1)$)
are the imaginary parts of the electron density correlation function 
and electron-phonon correlation function of the system in the magnetic field.

The $\Pi_2({\bf q}_{\|}, \Omega)$ function can be expressed 
in the Landau representation:\cite{Ting}
\begin{eqnarray}
&&\hspace{-0.7cm}\Pi _2({\bf q}_{\|},\Omega ) =  \frac 1{2\pi
l_{\rm B}^2}\sum_{n,n'}C_{n,n'}(l_{\rm B}^2q_{\|}^2/2) 
\Pi _2(n,n',\Omega),
\label{pi_2}\\
&&\hspace{-0.7cm}\Pi _2(n,n',\Omega)=-\frac2\pi \int d\varepsilon
\left [ f(\varepsilon )- f(\varepsilon +\Omega)\right ]\nonumber\\
&&\,\hspace{2cm}\times\,\,{\rm Im}G_n(\varepsilon +\Omega){\rm Im}G_{n'}(\varepsilon ),
\end{eqnarray}
where
$
C_{n,n+l}(Y)\equiv n![(n+l)!]^{-1}Y^le^{-Y}[L_n^l(Y)]^2
$
with $L_n^l(Y)$ the associate Laguerre polynomial, 
 $l_{\rm B}\equiv\sqrt{1/|eB|}$, $f(\varepsilon
)=\{\exp [(\varepsilon -\mu)/T_{\rm e}]+1\}^{-1}$ is the Fermi 
function at electron temperature $T_{\rm e}$. 
The density of states of the $n$-th Landau level
is modeled with a Gaussian form:\cite{Ando}
\begin{equation}
{\rm Im}G_n(\varepsilon)=-(\sqrt{2\pi}/\Gamma)
\exp[-{2(\varepsilon-\varepsilon_n)^2}/{\Gamma^2}]. \label{gauss}
\end{equation}
having a half-width
$
\Gamma=(8e\omega_c\alpha/\pi m \mu_0)^{1/2}
$
around the level center $\varepsilon_n$. Here 
$\mu_0$ is the linear mobility at lattice temperature $T$ 
in the absence of magnetic field, $\omega_c=eB/m$ 
and $\alpha$ is a semiempirical broadening parameter.

For time-independent ${\bf v}_0$,   
we obtain the transverse and longitudinal dc resistivities from Eq.\,(\ref{eqv0}):
 $R_{xy}=B/N_{\rm e}e$, 
\begin{equation}
R_{xx}=-{\bf F}_0\cdot{\bf v}_0/({N_{\rm e}^2e^2v_0^2}).
\end{equation}
The (linear) magnetoresistivity is its ${\bf v}_0\rightarrow 0$ limit. 
Within certain field range, $R_{xx}$ 
can be negative at small $v_0$, but increases with increasing $v_0$ 
and passes through zero at a finite $v_0$,\cite{Lei03} implying that 
the time-independent small-current solution is unstable and a spatially nonuniform\cite{Andreev} or
a time-dependent solution\cite{Ng} may develop, which exhibits measured zero resistance.
Therefore we identify the region where a negative dissipative magnetoresistance
develops as that of the ZRS. 

Assume that the 2DEG is contained in a thin sample suspended in vacuum 
at plane $z=0$.  
When an electromagnetic wave illuminates the plane perpendicularly 
with the incident electric field 
${\bf E}_{\rm i}(t)={\bf E}_{{\rm i}s}\sin(\omega t)+ {\bf E}_{{\rm i}c}\cos(\omega t)$, 
the HF electric field in the 2DEG  is   
\begin{equation}
{\bf E}(t)=\frac{N_{\rm e}e\,{\bf v}(t)}{2\epsilon_0 c}+{\bf E}_{\rm i}(t).
\end{equation}
Using this ${\bf E}(t)$ in Eqs.\,(\ref{eqv1}) and (\ref{eqv2}),
${\bf v}_1$ and ${\bf v}_2$ are explicitly expressed in terms of incident
field ${\bf E}_{{\rm i}s}$ and ${\bf E}_{{\rm i}c}$.

Note that 
all orders of real ($|n|>0$) and virtual ($n=0$) photon processes are
included in the summations over $n$ in Eqs.\,(\ref{eqf0}), (\ref{eqsp}) and (\ref{eqw}).
The phonon contributions to ${\bf F}_0$ and $S_{\rm p}$ 
have been neglected because of the low temperature setup in the experiments.
The short-range scatterers are considered to give 
the dominant contribution to the resistance and energy absorption
for the ultra-clean samples used.\cite{Lei05}
The numerical calculations are performed for $x$-direction 
(parallel to ${\bf E}_0$) linearly polarized incident microwave fields  
[${\bf E}_{{\rm i}s}=(E_{{\rm i}},0), {\bf E}_{{\rm i}c}=0$], using  
the material parameters of GaAs.\cite{Lei851}

\begin{figure}
\includegraphics [width=0.45\textwidth,clip=on] {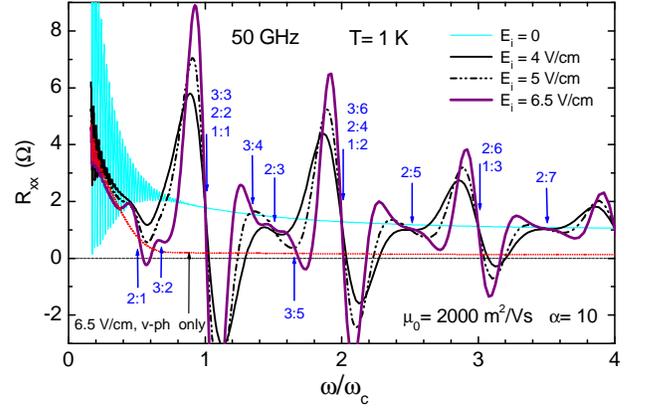}
\vspace*{-0.2cm}
\caption{The magnetoresistivity $R_{xx}$ of a GaAs-based 2DEG 
with $N_{\rm e}=3.0\times 10^{15}$\,m$^{-2}$, $\mu_0=2000$\,m$^2$/Vs and $\alpha=10$, 
subjected to 50\,GHz radiations of incident amplitudes $E_{{\rm i}s}=4,5,6.5$ and 
$8.5$\,V/cm at lattice temperature $T=1$\,K.}
\label{fig1}
\end{figure}
\begin{figure}
\includegraphics [width=0.45\textwidth,clip=on] {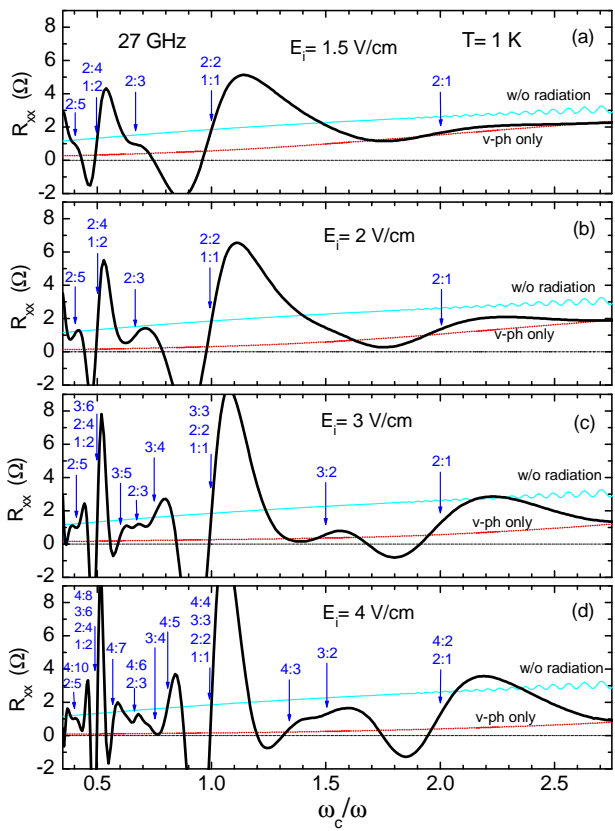}
\vspace*{-0.2cm}
\caption{The Magnetoresistivity $R_{xx}$ of a GaAs-based 2DEG 
with $N_{\rm e}=3.0\times 10^{15}$\,m$^{-2}$, $\mu_0=2000$\,m$^2$/Vs and $\alpha=5$, 
subjected to 27\,GHz radiations of incident amplitudes $E_{{\rm i}s}=1.5,2,3$ and 
$4$\,V/cm at lattice temperature $T=1$\,K.}
\label{fig2}
\end{figure}

Figure 1 shows the calculated magnetoresistivity $R_{xx}$ 
as a function of $\omega/\omega_c$ for a GaAs-based 2D system 
having electron density 
$N_{\rm e}=3.0\times 10^{15}$\,m$^{-2}$, 
linear mobility $\mu_0=2000$\,m$^2$/Vs and broadening parameter 
$\alpha=10$, irradiated by $50$\,GHz microwaves of  
three incident amplitudes $E_{{\rm i}}=4,5$ and 6.5\,V/cm 
at lattice temperature $T=1$\,K. We see that in addition to the
main maximum-minimum pairs at $\omega/\omega_c=1,2,3$ and 4,
prominent peak-valley pairs show up around $\omega/\omega_c=1/2,2/3,3/5,2/5$ and 2/7,
and the minima of pairs $\omega/\omega_c=1/2$ and $\omega/\omega_c=3/5$ drop down to
negative. 

Careful analysis shows that the appearance of oscillatory magnetoresistance comes from 
real photon-assisted electron transitions between different 
Landau levels as indicated in the summation of
the electron density-correlation function in Eq.\,(\ref{pi_2}).
We denote a real-photon assisted process in which an electron jumps across $l$ 
Landau-level spacings
with the assistance (emission or absorption) of $n$ photons as $n\omega$:$l\omega_c$, 
or $n$:$l$. This process contributes, 
in the $R_{xx}$ vs $\omega_c/\omega$ curve,
a pair structure consisting of a minimum and a maximum on both sides of $\omega_c/\omega=n/l$.
The location of its minimum or maximum may change somewhat depending on the strength 
of the incident microwave, but the node point, which is roughly in the center, 
essentially keeps at the position $\omega_c/\omega=n/l$.
Therefore we use its node position, rather than its minimum or the maximum,
to identify a pair structure.   
Thus, the single-phonon process $1$:$1$, the two-phonon process
$2$:$2$, the three-photon process $3$:$3$,$\cdots$, 
all contribute to the minimum--maximum pair around $\omega_c/\omega=1$;
the single-phonon process $1$:$2$, the two-phonon process
$2$:$4$, the three-photon process $3$:$6$,$\cdots$, 
all contribute to the minimum--maximum pair around $\omega_c/\omega=1/2$; etc.
These are indicated in Fig.\,1, as well as in Fig.\,2, where we plot  
the evolution of magnetoresistivity $R_{xx}$ with 
increasing radiation intensity $E_{{\rm i}}=1.5,2,3$ and 4\,V/cm  
vs $\omega_c/\omega$, for a GaAs-based system 
with $N_{\rm e}=3.0\times 10^{15}$\,m$^{-2}$, 
$\mu_0=2000$\,m$^2$/Vs and $\alpha=5$, irradiated by $27$\,GHz microwaves.
Several other valley-peak pairs associated with 2-, 3-, and 4-photon assisted processes
are also identified. 

The predicted two-photon structure $2$:$1$ centered at $\omega_c/\omega=2$ with  
a minimum around $\omega_c/\omega\approx 1.73$ (Fig.\,2(a) and (b)) 
reasonably compares with the second harmonic minimum around $\omega_c/\omega=8/5$ 
observed in Refs.\,\onlinecite{Mani,Mani04}. The two-photon structures 
$2$:$3$ centered at $\omega_c/\omega=2/3$
and $2$:$5$ centered at $\omega_c/\omega=2/5$ are 
in agreement with experimental findings.\cite{Mani,Mani04,Zud03,Zud04} 

Note that, with the progressive emergence of new multiphoton-related pairs
when increasing radiation power, the peaks (valleys) of the low-order photon 
related pairs become narrower, as is clearly seen 
for the single-photon related pairs at $\omega/\omega_c=1,2$ and 3
in both figures. This feature and the anticipated positions of the peaks and 
possible ZRSs, agree with the recent experimental observation.\cite{Zud05}

Another interesting aspect is that, concomitantly with enhanced $R_{xx}$ oscillation, 
the average magnetoresistance descends down significantly
with increasing radiation power. This resistance drop is due to effect of virtual photon 
processes, i.e. intra-Landau-level electron scattering by impurities 
with simultaneous emission and absorption of an arbitrary number of photons. 
To show this we plot the resistivity contributed 
from the virtual photon processes alone, i.e. the $n=0$ term ($J_0$) 
in Eq.\,(\ref{eqf0}), in both figures. The resistance suppression
appears almost throughout the whole magnetic field range. It may not be previously 
noticed in the region where exhibits strong $R_{xx}$ oscillation and ZRS.       

This work was supported by Projects of the National Science Foundation of China
and the Shanghai Municipal Commission of Science and Technology.


\begin{thebibliography}{99}
\bibitem{Zud01} M. A. Zudov, R. R. Du, J. A. Simmons, and J. L. Reno, Phys. Rev. B
{\bf 64}, 201311(R) (2001).

\bibitem{Ye} P. D. Ye, L. W. Engel, D. C. Tsui, J. A. Simmons, J. R. Wendt, G. A. Vawter, 
and J. L. Reno, Appl. Phys. Lett. {\bf 79}, 2193 (2001).

\bibitem{Mani} R. G. Mani, J. H. Smet, K. von Klitzing, V. Narayanamurti, 
W. B. Johnson, and V. Umansky, Nature {\bf 420}, 646 (2002).

\bibitem{Zud03}  M. A. Zudov, R. R. Du, L.N. Pfeiffer, and K. W. West, Phys. Rev. Lett.
{\bf 90}, 046807 (2003).

\bibitem{Yang} C. L. Yang, M. A. Zudov, T. A. Knuuttila, R. R. Du, L. N. Pfeiffer,
and K. W. West, Phys. Rev. Lett. {\bf 91}, 096803 (2003).

\bibitem{Dor03} S. I. Dorozhkin, JETP Lett. {\bf 77}, 577 (2003).

\bibitem{Mani04} R. G. Mani, J. H. Smet, K. von Klitzing, V. Narayanamurti,
W. B. Johnson, and V. Umansky, Phys. Rev. Lett. {\bf 92}, 146801 (2004).

\bibitem{Will} R. L. Willett, L. N. Pfeiffer, and K. W. West, Phys. Rev. Lett. {\bf 93}, 
026804 (2004).

\bibitem{Kovalev} A. E. Kovalev, S. A. Zvyagin, C. R. Bowers, J. L. Reno, 
 and J. A. Simmons, Solid State Commun. {\bf 130}, 379 (2004).

\bibitem{Mani-apl} R. G. Mani, Physica E {\bf 25}, 189 (2004); Appl. Phys. Lett. 
{\bf 85}, 4962 (2004).

\bibitem{Zud04} M. A. Zudov, Phys. Rev. B {\bf 69}, 041304(R) (2004).

\bibitem{Stud} S. A. Studenikin, M. Potenski, A. Sachrajda, M. Hilke, L. N. Pfeiffer, 
and K. W. West, IEEE Tran. on Nanotech. {\bf 4}, 124 (2005).

\bibitem{Dor05} S. I. Dorozhkin, J. H. Smet, V. Umansky, K. von Klitzing, Phys. Rev. B
{\bf 71}, 201306(R) (2005).

\bibitem{Zud05} M. A. Zudov, R. R. Du, L. N. Pfeiffer, and K. W. West, Phys. Rev. B
{\bf 73}, 041303(R) (2006).

\bibitem{Ryz} V. I. Ryzhii, Sov. Phys. Solid State {\bf 11}, 2087 (1970).

\bibitem{Ryz86}  V. I. Ryzhii, R. A. Suris, and B. S. Shchamkhalova, Sov. Phys.-Semicond.
{\bf 20}, 1299 (1986).

\bibitem{Anderson} P. W. Anderson and W. F. Brinkman,  cond-mat/0302129.

\bibitem{Koul} A. A. Koulakov and M. E. Raikh, Phys. Rev. B {\bf 68}, 115324 (2003).

\bibitem{Andreev} A. V. Andreev, I. L. Aleiner, and A. J. Mills, Phys. Rev. Lett. {\bf 91}, 
056803 (2003).

\bibitem{Durst} A. C. Durst, S. Sachdev, N. Read, and S. M. Girvin, Phys. Rev. Lett. 
{\bf 91}, 086803 (2003).

\bibitem{Xie} J. Shi and X. C. Xie, Phys. Rev. Lett. {\bf 91}, 086801 (2003).

\bibitem{Dmitriev} I. A. Dmitriev, A. D. Mirlin, and D. G. Polyakov, Phys. Rev. Lett. 
{\bf 91}, 226802 (2003). 

\bibitem{Lei03} X. L. Lei and S. Y. Liu, Phys. Rev. Lett. {\bf 91}, 226805 (2003);
 X. L. Lei, J. Phys.: Condens. Matter {\bf 16}, 4045 (2004). 

\bibitem{Ryz05199} V. Ryzhii and R. Suris,  J. Phys.: Condens. Matter {\bf 15}, 
6855 (2003).

\bibitem{Vav} M. G. Vavilov and I. L. Aleiner, Phys. Rev. B {\bf 69}, 035303 (2004).

\bibitem{Mikh} S. A. Mikhailov, Phys. Rev. B {\bf 70}, 165311 (2004).

\bibitem{Dietel} J. Dietel, L. Glazman, F. Hekking, F. von Oppen, Phys. Rev. B {\bf 71}, 
045329 (2005).

\bibitem{Torres} M. Torres and A. Kunold, Phys. Rev. B {\bf 71}, 115313 (2005).

\bibitem{Dmitriev04} I. A. Dmitriev, M. G. Vavilov, I. L. Aleiner, A. D. Mirlin,
and D. G. Polyakov, Phys. Rev. B {\bf 71}, 115316 (2005).

\bibitem{Inar} J. I\~{n}arrea and G. Platero, Phys. Rev. Lett. {\bf 94}, 016806 (2005).

\bibitem{Lei05} X. L. Lei and S. Y. Liu, Phys. Rev. B {\bf 72}, 075345 (2005);
Appl. Phys. Lett. {\bf 86}, 262101 (2005).

\bibitem{Ryz05}  V. Ryzhii, Jpn. J. Appl. Phys. {\bf 44}, 6600 (2005).

\bibitem{Joas05} C. J. Joas, J. Dietel, and F. von Oppen, Phys. Rev. B {\bf 72}, 
165323 (2005).

\bibitem{Ng} T. K. Ng  and  Lixin Dai, Phys. Rev. B {\bf 72}, 235333 (2005).

\bibitem{Lei85} X. L. Lei and C. S. Ting, Phys. Rev. B {\bf 32}, 1112 (1985);
X. L. Lei, J. Appl. Phys. {\bf 84}, 1396 (1998).

\bibitem{Liu} S. Y. Liu and X. L. Lei, J. Phys.: Condens. Matter {\bf 15}, 4411 (2003).

\bibitem{Ting}  C. S. Ting, S. C. Ying, and J. J. Quinn,
Phys. Rev. B {\bf 14}, 5394 (1977).

\bibitem{Ando} T. Ando, A. B. Fowler, and F. Stern, Rev. Mod. Phys. {\bf 54},
437 (1982).

\bibitem{Lei851} X. L. Lei, J. L. Birman, and C. S. Ting, J. Appl. Phys. 
{\bf 58}, 2270 (1985).


\end{thebibliography}
\end{document}